\documentclass[prb,twocolumn,showpacs,preprintnumbers,amsmath,amssymb]{revtex4}

\usepackage{graphicx}
\usepackage{epsfig}
\usepackage{dcolumn}
\usepackage{bm}
\usepackage{placeins}
\usepackage{flafter}
\usepackage{color}
\definecolor{royal}{rgb}{0,0,.60}

\newcommand{\ketbr}{$\kappa$-(BEDT-TTF)$_2$Cu[N(CN)$_2$]Br}

\begin{document}

\title{Influence of the cooling rate on low-temperature Raman and infrared reflection spectra of partially deuterated \ketbr}

\author{M. Maksimuk}
\altaffiliation{On leave from the Institute of Solid State Physics, RAS, Chernogolovka 142432 Russia}
\email{maksimuk@issp.ac.ru}
\author{K. Yakushi}
\affiliation{Institute for Molecular Science, Myodaiji, Okazaki, Aichi 444-8585}
\author{H. Taniguchi}
\author{K. Kanoda}
\affiliation{Department of Applied Physics, University of Tokyo, Hongo, Tokyo, 113-0033}
\author{A. Kawamoto}
\affiliation{Department of Physics, University of Hokkaido, Sapporo, Hokkaido 060-0810}

\date{\today}

\begin{abstract}
Raman and IR spectra of \ketbr\ [BEDT-TTF denotes bis(ethylenedithiolo)tetrathiafulvalene] and its deuterated and partially deuterated analogues were measured at temperatures between 5 and 300 K and cooling rates from 1 to 20 K/min. It was found that, in partially deuterated samples, the interdimer electron-molecular vibration splitting of $\nu_3$ mode in Raman spectra, the magnitude of the resonance enhancement in Raman spectra, and linewidths of some phonon peaks both in Raman and infrared spectra depend on the cooling rate. These observations were explained by disorder-related effects. 
\end{abstract}

\maketitle

\section{Introduction}

\ketbr\ [BEDT-TTF denotes bis(ethylenedithiolo)tetrathiafulvalene] is one of the most well-known organic superconductors.\cite{lang} Still, a number of issues related to its physical properties remain unclear, including the nature the superconducting state. This compound lies on the border between metals and insulators and, below 50 K, regions of an antiferromagnetic insulating phase appear inside its mostly metallic volume. Upon deuteration, the amount of the insulating phase gradually increases with increasing deuterium content.\cite{tani} There exists a set of progressively deuterated isotopic analogues of \ketbr, which are denoted as d[n,n] with n = 0, 1, 2, 3, or 4 being the number of deuterium atoms at both ends of a BEDT-TTF molecule.\cite{kawa1} The last member of this sequence, d[4,4], is mostly insulating at low temperatures. The cooling rate is another important factor affecting the phase separation: the faster the cooling the larger is the amount of the insulating phase.\cite{tani} It is believed that influence of the cooling rate on phase separation is related to ordering of ethylene groups of BEDT-TTF molecules, taking place approximately from 140 down to 50 K as a glass transition.\cite{saito, akutsu, stalcup} Fast cooling freezes the disorder in ethylene groups, more effectively in heavier deuterated ones. The mechanism of the effect of ethylene-group disorder on phase separation is not understood in detail at the moment. A disorder of ethylene groups probably results in disorder of transfer integrals.\cite{muller} Yoneyama \textit{et al.}\cite{yone} suggested that the disorder in ethylene groups decreases inhomogeneously upon cooling, with the formation of domains having some amount of disorder and domains that are free of disorder. The former then become insulating and the latter become metallic.

Most of experimental studies carried out so far deal with d[0,0]. d[4,4] is less investigated, and few works are devoted to partially deuterated crystals. At the same time, the superconductiong/insulating phase transition in d[2,2] and d[3,3] can be easier controlled by cooling rate.\cite{tani}

In our previous publication,\cite{maksimuk} we reported polarised Raman spectra of \ketbr\ along with its several isotopic analogues and discussed the splittings of $\nu_2$ and $\nu_3$ molecular modes. Here, we present the cooling-rate and temperature dependences of Raman and IR spectra of d[0,0], d[2,2], d[3,3], and d[4,4] single crystals in the frequency range corresponding to C=C stretching modes. We supposed that the cooling rate should have significant influence on the phonon spectra; however, only rather small effects were observed. Below we report and discuss these findings.

\section{Experimental}

The experimental procedure was the same as in Ref. \onlinecite{maksimuk}. The crystals were grown by a standard electrochemical oxidation. The Raman spectra were measured in the back-scattering geometry by a Renishaw Ramascope System-1000 spectrometer. A 514-nm argon laser was used for excitation. The laser power on the sample surface was $\approx$ 50 $\mu$W, the spot diameter being $\approx$ 5 $\mu$m. The typical time of one measurement was 15 min. The experimental error of a wavenumber absolute value was 1 cm$^{-1}$, the error of relative wavenumber shift was 0.1 cm$^{-1}$, the width of the instrument response function was 3.3 cm$^{-1}$.

The polarised reflection spectra in 600--4000 cm$^{-1}$ region were obtained with a FT-IR Nicolet Magna 760 spectrometer combined with a Spectratech IR-Plan microscope. The spectral resolution was 4 cm$^{-1}$.

 Samples were held inside a helium-flow Oxford CF1104s cryostat. The sample was fixed to the holder by a small amount of the silicone grease. The temperature was monitored by a silicon diode sensor attached inside the holder just below the sample crystal.

To obtain temperature dependences the following number of different samples shown in parentheses were used: for IR measurements, d[0,0] (1), d[2,2] (1), d[3,3] (1); for Raman measurements, laser polarisation $\mathbf{c}$, d[0,0] (2), d[2,2] (2), d[3,3] (3), d[4,4] (1); for Raman measurements, laser polarisation $\mathbf{b}$, d[0,0] (2), d[2,2] (1), d[3,3] (1), d[4,4] (1). For Raman measurements we used different samples in $\mathbf{c}$ and $\mathbf{b}$ polarisations. Several more samples of each sort were used for a rough check of the sample-to-sample spectra variations.

To obtain the parameters of separate lines, the least-square fitting by Lorentz functions was used. For each mode $\nu$ the position $\omega(\nu)$, the width $\delta(\nu)$ and the integrated intensity $I(\nu)$ were extracted. Below we will somewhere designate the laser polarisation and the type of the crystal in parentheses; for example, $\omega(\nu_2,\bm{b},2)$ means the position of $\nu_2$ mode for d[2,2] crystal and $\mathbf{b}$ laser polarisation. The polarisation of the scattered light was not analysed, so only the polarisation of the laser radiation is indicated.

At any temperature, all spectral lines could be fitted quite well except their wings. When a weak line is located near an intense one, the width of the former is influenced by the latter. For that reason, we used different laser polarisation for different modes to obtain temperature dependences. 

\section{Results and discussion}

\begin{figure}[h]
\includegraphics[width=0.5\textwidth,height=0.5\textwidth,keepaspectratio=true]{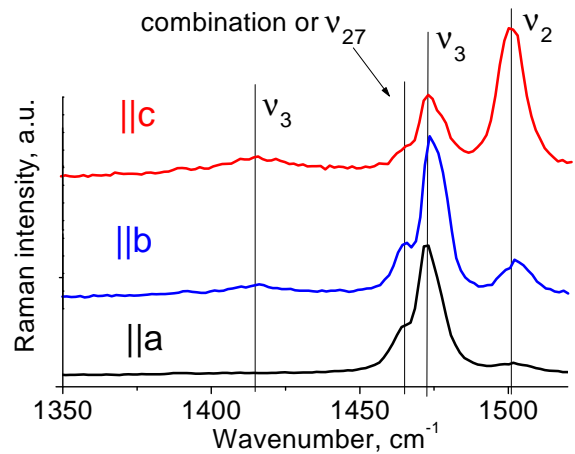}
\caption{\label{fig:fig1} Raman lines of C=C stretching vibrations for d[0,0] at 5 K.}
\end{figure}

In Fig.~\ref{fig:fig1}, the Raman spectra of d[0,0] in the region of C=C stretching vibrations are shown for different polarisations of the laser. The $\nu_3$ molecular mode gives rise to two components: $\nu_3(A_g+B_{1g})$ near 1475 cm$^{-1}$ and $\nu_3(B_{2g}+B_{3g})$ near 1415 cm$^{-1}$; the splitting between them is determined by the dimer--dimer electron molecular vibration (EMV)  interaction.\cite{maksimuk} The line originating from the mode $\nu_2$ appears near 1500 cm$^{-1}$. The line near 1465 cm$^{-1}$ was identified by us as $\nu_{27}$ or a combination.\cite{maksimuk}

We found certain cooling-rate dependences for d[2,2] and d[3,3]. For d[0,0] and d[4,4] no dependences were found. We used "fast" and "slow" cooling procedures. "Fast" cooling means approximately 20 K/min which was at the limit of the equipment used. We tried different "slow" cooling procedures: (I) approximately 0.3--1 K/min cooling in some temperature range between 50 and 150 K, (II) annealing at some constant temperature between 60 and 80 K, (III) some combination of 0.3--1 K/min cooling and annealing.

\begin{figure}[h]
\includegraphics[width=0.5\textwidth,height=0.5\textwidth,keepaspectratio=true]{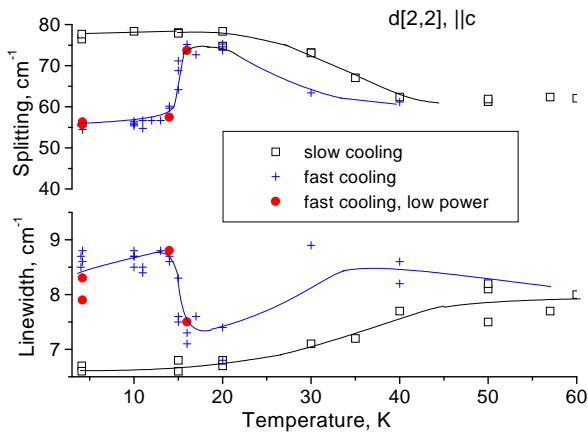}
\caption{\label{fig:fig2} Top, temperature dependence of $\nu_3$ dimer--dimer splitting $\omega(\nu_3(A_g+B_{1g}))-\omega(\nu_3(B_{2g}+B_{3g}))$. Bottom, temperature dependence of $\nu_3(A_g+B_{1g})$ linewidth. All data are taken for the same d[2,2] crystal. Laser polarisation is $\mathbf{c}$. Squares '$\Box$' show slow cooling (see text); crosses '\textcolor{blue}{+}' show fast cooling, 50 $\mu$W laser power; circles '{\large\textcolor{red}{$\bullet$}}' show fast cooling, 15 $\mu$W laser power. Curves are a guide to the eye.}
\end{figure}

The results obtained with d[2,2] crystals showed strong variation from sample to sample. In some of the samples we could see cooling rate dependence, in others we could not. The results depended also on the way the sample was fixed. As a rule, simply after removing the sample from the holder, cleaning it from silicone grease by a solvent, and reattaching it by another small portion of grease, results different from initial ones were obtained. During the cooling the samples could jump or rotate a little if the amount of the grease was was insufficient. By this reason we could not totally remove the influence of grease. However, if the sample position remained stable, the results for this particular sample were reproducible from cooling to cooling. At low temperatures, $\nu_2$ and $\nu_3$ lines looked additionally split. We can not characterise the cooling rate dependence for d[2,2] in general. However, in Fig.~\ref{fig:fig2} we present the results for one specific d[2,2] crystal with the biggest difference between "slow" and "fast" cooling --- first, as an example, and second, because we used these data to estimate the overheating of the sample in the laser spot. 

For Fig.~\ref{fig:fig2} "slow" cooling sequence was the following: cooling to 75 K, 1 hour annealing, 0.3 K/min cooling to 70 K, 4 hours annealing, 0.3 K/min cooling to 60 K, 1 hour annealing, 0.3 K/min cooling to 50 K, and cooling to 5 K. (The results obtained are shown by squares.) Quite different dependence was obtained upon  "fast" cooling, with most striking difference observed between 10 and 20 K; being rather unusual these results were, however, reproducible from one cooling cycle to another. To check the laser overheating we performed these measurements upon "fast" cooling under two levels of laser power: 15 $\mu$W (shown by dark circles in Fig.~\ref{fig:fig2}, accumulation time 30 min), and 50 $\mu$W (shown by crosses, accumulation time 15 min). 

As one can see in Fig.~\ref{fig:fig2}, the data for the linewidth and the splitting do not depend on the laser power allowing us to estimate that, near 10--20 K, the overheating did not exceed 3 K. We also tried to measure the overheating by another method: we increased laser power 3 times (up to about 150 $\mu$W), measured the signal at 90 K and estimated the overheating by change of the width and the position of $\nu_2$, $\nu_3$, and 1465-cm$^{-1}$ line. Estimations based on different parameters gave different values of the overheating, however, all being between 10 and 50 K (i. e., 3--15 K for our normal laser power). 

We cannot state for sure that, at nominal 5 K, the crystals were really below the critical temperature of the superconducting transition. Several measurements with d[3,3] and d[2,2] crystals at 5 K upon "fast" cooling were carried out at lowered laser intensity, and no essential changes in any spectral parameter in comparison to the case of normal laser intensity were observed; however, we have very little statistics for such measurements.

d[0,0], d[3,3], d[4,4] crystals showed no evident variation from sample to sample. We attached several d[3,3] samples to the holder, and these samples underwent the same cooling procedure. The results were very close. Thus we can speak about the behaviour of d[3,3] after "fast" or "slow" cooling in general.

It is worth mentioning that we did not see any considerable difference between different places of the same sample in any of d[0,0], d[2,2], d[3,3], d[4,4] crystals.

\begin{figure}[h]
\includegraphics[width=0.5\textwidth,height=0.5\textwidth,keepaspectratio=true]{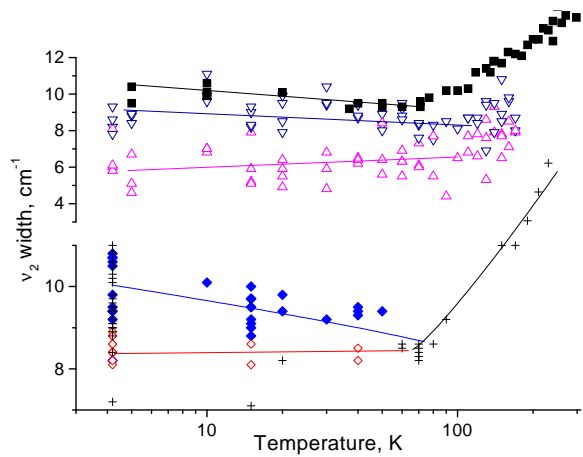}
\caption{\label{fig:fig3} Temperature dependence of $\delta(\nu_2,\bm{c})$. d[0,0], the low-frequency component (up triangles, \textcolor{magenta}{$\triangle$}); d[0,0], the high-frequency component (down triangles, \textcolor{royal}{$\bigtriangledown$}; d[4,4] (squares, $\blacksquare$); d[3,3] (solid diamonds, \textcolor{blue}{$\blacklozenge$}, for the fast cooling; open diamonds, \textcolor{red}{$\diamondsuit$}, for the slow cooling; crosses, +, for the unfinished slow cooling). Lines are a guide to the eye.}
\end{figure}

\begin{figure}[h]
\includegraphics[width=0.5\textwidth,height=0.5\textwidth,keepaspectratio=true]{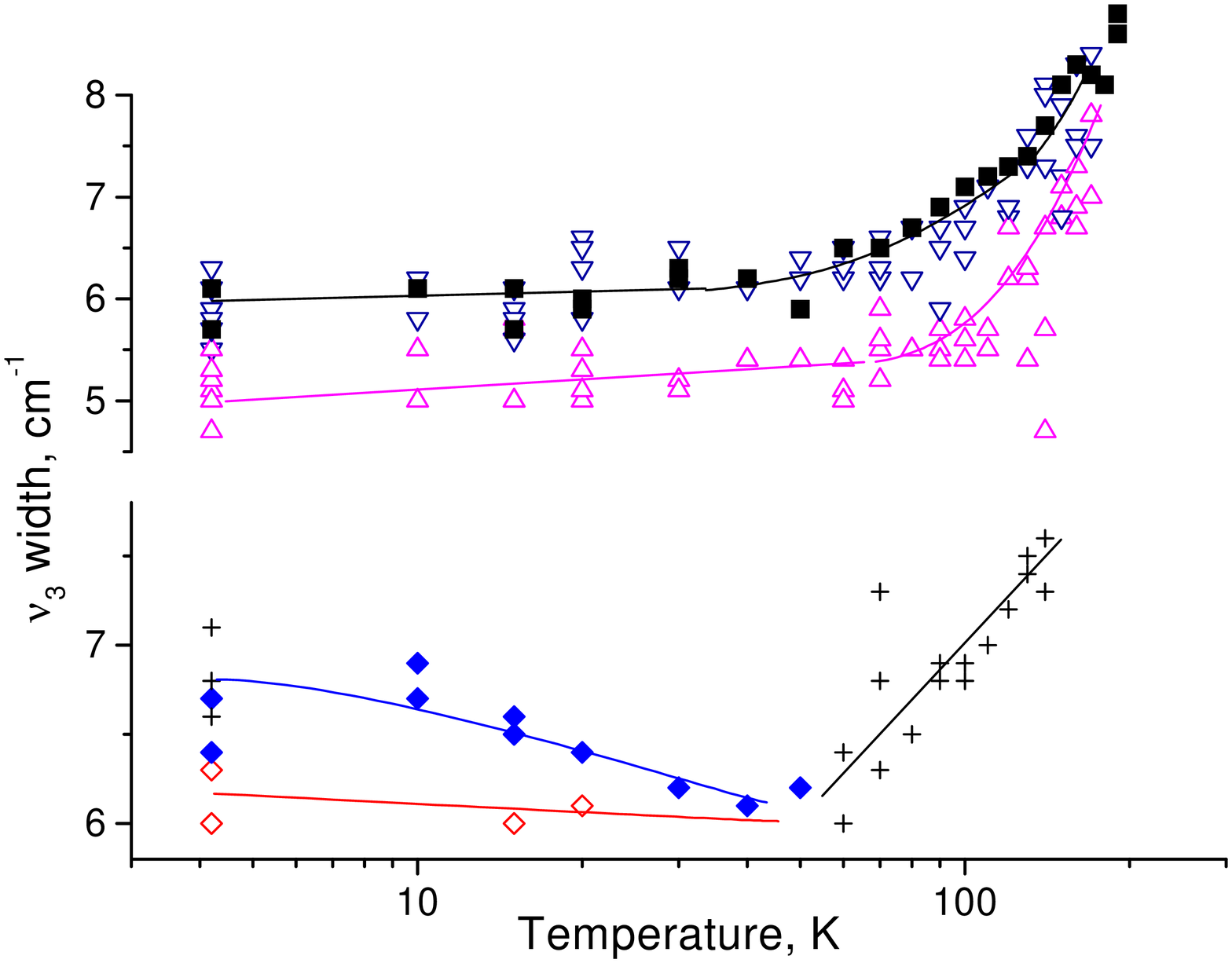}
\caption{\label{fig:fig4} Temperature dependence of  $\delta(\nu_3(A_g+B_{1g}),\bm{b})$. d[0,0], the low-frequency component (up triangles, \textcolor{magenta}{$\triangle$}); d[0,0], the high-frequency component (down triangles, \textcolor{royal}{$\bigtriangledown$}; d[4,4] (squares, $\blacksquare$); d[3,3] (solid diamonds, \textcolor{blue}{$\blacklozenge$}, for the fast cooling; open diamonds, \textcolor{red}{$\diamondsuit$}, for the slow cooling; crosses, +, for the unfinished slow cooling). Lines are a guide to the eye.}
\end{figure}

The temperature dependences of $\delta(\nu_2,\bm{c})$ and $\delta(\nu_3(A_g+B_{1g}),\bm{b})$ are shown in Fig.~\ref{fig:fig3} and Fig.~\ref{fig:fig4}, respectively. For d[0,0] crystals $\nu_2$, $\nu_3$ modes are split.\cite{maksimuk} We designate the components of these doublets as $\nu_2^l$, $\nu_2^h$, $\nu_3^l$, $\nu_3^h$. The linewidths of all components are shown in upper panels of Fig.~\ref{fig:fig3} and ~\ref{fig:fig4} together with $\delta(\nu_2,4), \delta(\nu_3, 4)$. Below 20--40 K  $\delta(\nu_2^l,0)$ decreases with decreasing temperature but $\delta(\nu_2^h,0)$ increases.

Above 20--40 K the data for d[2,2], d[3,3], d[4,4] coincide, the linewidths increasing linearly with the increase of the temperature with the slope $(1.9\pm0.1)\times10^{-2} cm^{-1}K^{-1}$ for $\delta(\nu_3)$ and $(2.5\pm0.2)\times10^{-2} cm^{-1}K^{-1}$ for $\delta(\nu_2)$. At low temperature $\delta(\nu_2)$ is about 1--2 cm$^{-1}$ larger than $\delta(\nu_3)$. Below 20--40 K, the temperature dependences of the linewidth no longer coincide for the samples with different deuteration.

For d[3,3] the cooling rate influences the result. After "slow" cooling (open diamonds) the linewidth remains constant, after "fast" cooling it increases with decreasing temperature. Crosses in Figs.~\ref{fig:fig3} and ~\ref{fig:fig4} represent the data obtained when, to make a measurement in a region 50--150 K, we interrupted the "slow" cooling procedure. Above 150 K the cooling rate was not important and the data are shown by crosses too. 
\begin{figure}[h]
\includegraphics[width=0.5\textwidth,height=0.5\textwidth,keepaspectratio=true]{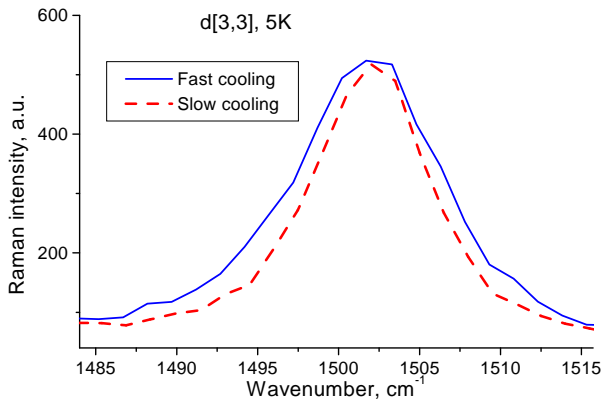}
\caption{\label{fig:fig5} Raman scattering of d[3.3] ($\nu_2$ mode) at 5 K after "slow" cooling (dashed line) and "fast" cooling (solid line). Polarisation $\bm{c}$.}
\end{figure}

\begin{figure}[h]
\includegraphics[width=0.5\textwidth,height=0.5\textwidth,keepaspectratio=true]{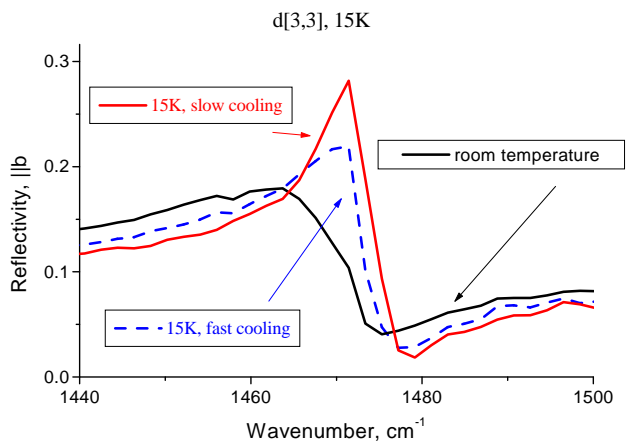}
\caption{\label{fig:fig6} Reflectivity of d[3,3] in $\bm{c}$ polarisation after "slow" and "fast" cooling.}
\end{figure}

In Fig.~\ref{fig:fig5} we show how the difference between "slow" and "fast" coolings looks like. As a result of the "fast" cooling, $\nu_2$ widens uniformly. A similar broadening effect can be seen in IR reflectivity spectra too, as it is shown in Fig.~\ref{fig:fig6}. In the latter case, the integrated intensity of the phonon line did not depend of the cooling rate.

An increase in the linewidth with decreasing temperature looks quite unusual in itself. It can be seen also in $\delta(\nu_2,\bm{c},4)$ and $\delta(\nu_2^h,\bm{c},0)$ (Fig.~\ref{fig:fig3}). The disorder in ethylene groups produced by fast cooling occurs approximately from 150 down to 50 K (Refs. \onlinecite{saito, akutsu}). It is also this temperature region that we found to influence our results for d[3,3].  The phase separation occurs below 80 K (Ref. \onlinecite{tani}). So it would appear reasonable to relate the low-temperature broadening of spectral lines and the cooling-rate dependence of the linewidth with the phase separation rather than the ethylene-group disorder itself. However, the broadening can not be produced just as a result of differences between metallic and insulating phase in $\omega(\nu_2)$ and $\omega(\nu_3)$ because the broadening can be seen both in almost totally metallic\cite{tani} d[0,0] and almost totally insulating\cite{tani} d[4,4]. 

There is a possibility to explain the broadening by increased scattering of phonons at the domain boundaries. The amount of superconducting phase is very different in d[0,0], d[2,2], d[3,3] and d[4,4], however, they all can have zero resistance below the temperature of the superconducting transition. So the bulk of the material probably consists of insulating domains embedded into the metallic matrix. It may be guessed that the characteristic size of insulating domains is the same for any sort of crystal. When the temperature decreases, the insulator--metal boundaries become more pronounced and can scatter phonons more effectively. Slow cooling increases the domain size in d[2,2], d[3,3], and probably in d[0,0], but, for unknown reason, not in d[4,4]. An upper limit for an insulating domain size $d$ can be estimated as $d\approx c\tau\approx c/\delta$, where $c$ is the sound velocity, $\tau$ is the phonon lifetime, $\delta$ is the phonon linewidth. It gives several nanometres which is quite consistent with the model proposed by Yoneyama \textit{et al.}\cite{yone}. 
\begin{figure}[h]
\includegraphics[width=0.5\textwidth,height=0.5\textwidth,keepaspectratio=true]{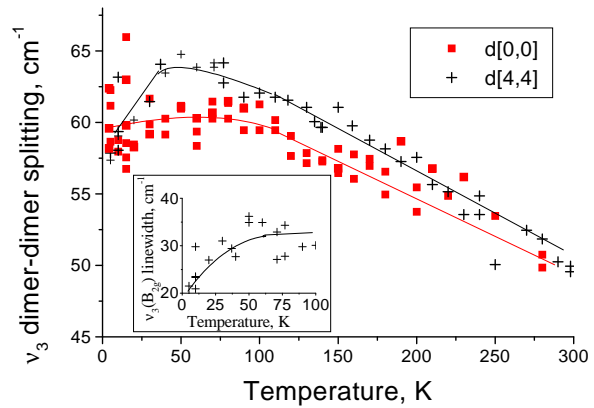}
\caption{\label{fig:fig7} Temperature dependence of the splitting between $\nu_3(B_{2g}, B_{3g})$ and $\nu_3(A_g, B_{1g})$ for d[0,0], d[4,4]. The inset shows the temperature dependence of $\nu_3(B_{2g}, B_{3g})$ linewidth. Lines are guide to the eye.}
\end{figure}

Another parameter that depends on the cooling rate is EMV dimer--dimer splitting  of $\nu_3$ mode, $\omega(\nu_3(A_g, B_{1g}))-\omega(\nu_3(B_{2g}, B_{3g}))$. In Fig.~\ref{fig:fig7}, we show its temperature dependence for d[0,0] and d[4,4]. Above 50 K the splitting decreases significantly (by 25\% per 200 K) with increasing temperature, and there is no considerable difference between mostly metallic d[0,0] and mostly insulating d[4,4]. For d[0,0] below 50 K the splitting remains constant or maybe decreases a little with decreasing temperature. For d[4,4] the splitting evidently decreases below 50 K, the linewidth $\delta(\nu_3(B_{2g}, B_{3g}))$ substantially decreasing too (see the inset). (For d[0,0], d[2,2], d[3,3] the values obtained for this linewidth were more scattered and no definite change could be detected.)

According to Eldridge \textit{et al.}\cite{eldr}, for d[0,0] the behaviour of the \textit{intradimer} EMV coupling is quite different. It continuously increases with increasing temperature but mainly below 100 K. For $\kappa$-(BEDT-TTF)$_2$Cu[N(CN)$_2$]Cl the temperature behaviour of the \textit{intradimer} EMV coupling is different again, the splitting decreasing below 50 K and increasing above 100 K with increasing temperature.\cite{korn}

For the temperature behaviour of the \textit{interdimer} EMV coupling we can consider two important factors. 
The first one is the temperature change of lattice parameters\cite{kund,toyota} which can influence \textit{interdimer} interaction more than \textit{intradimer} interaction. However, calculations (see Ref. \onlinecite{noga}, for example) show that (i) the \textit{intradimer} transfer integrals are influenced not less that \textit{interdimer} ones and (ii) about 1 \% change of the lattice parameters really taking place between 100 and 300 K leads to less than 10 \% change in the transfer integrals. Probably, another factor, the temperature variation of the antiferromagnetic correlations, can be important. The charge transfer and EMV coupling act predominantly between dimers with antiparallel spins. Below $\approx$300 K these correlations increase gradually with the decrease of temperature.\cite{nakaz} Only short-range correlations are important which can result in the absence of substantial difference between metallic d[0,0] and dielectric d[4,4]. 

The influence of the antiferromagnetic correlations on the EMV coupling can be roughly illustrated in the frame of Rice's dimer model.\cite{rice} If the spins are antiferromagnetically ordered then the dimer--dimer charge-transfer transition results in the singlet state with the effective dimer Coulomb energy $U_{eff}^{\uparrow\downarrow}=2t_{ra}+U/2-\sqrt{U^2/4+4t^2_{ra}}$ (Ref. \onlinecite{kino}), where $U$ is the intramolecular Coulomb energy and $ t_{ra}$ is the intradimer transfer integral. If the spins are completely disordered then, in the half of all cases, after the dimer--dimer charge-tranfer transition there occurs the triplet state with the effective dimer Coulomb energy $U_{eff}^{\uparrow\uparrow}=2t_{ra}$. The difference $U_{eff}^{\uparrow\uparrow}-U_{eff}^{\uparrow\downarrow}=\sqrt{U^2/4+4t^2_{ra}}-U/2\approx$0.15 eV (taking $U\approx$1  eV, $t_{ra}\approx$0.2 eV). In comparison with $U_{eff}\approx$0.6 eV, it is a considerable value but the estimate is very rough because the value for $U$ is not accurately known\cite{taji,park,mcken}. In addition, $U$ is affected strongly by the screening effect. We can expect that spin ordering leads to a decreased screening, thus increasing $U$ and enhancing the dependence of $U_{eff}$ upon spin--spin correlations.

However, there exists an opposite factor, namely, in the antiferromagnetic state as a result of the charge transfer an electron obtains an additional exchange energy, approximately estimated as $3*2t^2_{er}/U_{eff}\approx$0.1 eV (interdimer transfer integral $t_{er}\approx$0.1 eV). (As to the difference in the intermolecular Coulomb interaction inside a dimer between the singlet state and the triplet state, this difference is very small, less than 0.01 eV (Ref.  \onlinecite{fort})). 

The above-estimated values have opposite signs, and we cannot assess their sum even roughly.However, in any case the antiferromagnetic ordering can essentially influence $U_{eff}$ and, therefore, the EMV coupling. 

Below 50 K the two-dimensional antiferromagnetic short-range spin--spin correlations are already almost as large as possible, both in insulating and metallic phases, increasing no longer. However, in the antiferromagnetic phase screening can continue to decrease with decreasing temperature that resuls to an increase of the intramolecular Coulomb energy $U$ and therefore to a small increase of $U_{eff}$ and to a small decrease of  the EMV coupling. In such a case the decrease of $\delta(\nu_3(B_{2g}, B_{3g}))$ for d[4,4] can be explained as the result of decreasing amount of the metallic phase in the bulk of the crystal.
\begin{figure}[h]
\includegraphics[width=0.5\textwidth,height=0.5\textwidth,keepaspectratio=true]{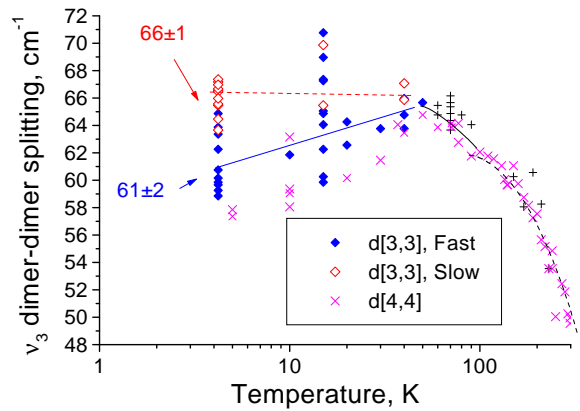}
\caption{\label{fig:fig8} Temperature dependence of the dimer--dimer EMV splitting of $\nu_3$ for d[3.3] after "slow" cooling (open diamonds, \textcolor{red}{$\diamondsuit$}), "fast" cooling (solid diamonds, \textcolor{blue}{$\blacklozenge$}), and above 50 K (crosses, +). The results for d[4,4] are also shown (crosses, \textcolor{magenta}{$\times$}). Lines are a guide to the eye. Polarization $\bm{c}$.}
\end{figure}

For d[2,2] and d[3,3] the dimer--dimer EMV splitting depends on the cooling rate. Fig.~\ref{fig:fig8} demonstrates this effect for d[3,3]. Below 80 K after "slow" cooling the splitting is approximately constant (open diamonds on Fig.~\ref{fig:fig8}), after "fast" cooling (solid diamonds) it decreases with decreasing temperature. At 5 K we estimate the splitting for d[3,3] as 66$\pm$1 cm$^{-1}$ after "slow" cooling and 61$\pm$2 cm$^{-1}$ after "fast" cooling. For comparison, we also present in Fig.~\ref{fig:fig8} the data for d[4,4] (crosses). After "fast" cooling the behaviour of d[3,3] looks similar to d[4,4].

Concerning the intensities of Raman signal, for all the samples under the 514-nm laser irradiation, the ratio $I(\nu_2,\bm{b})/I(\nu_3,\bm{b})$ is temperature-independent, but the ratio $I(\nu_2,\bm{c})/I(\nu_3,\bm{c})$ increases linearly with decreasing temperature. To a first approximation the increase in $I(\nu_2,\bm{c})/I(\nu_3,\bm{c})$ occurs purely at the expense of $I(\nu_2,\bm{c})$ increasing linearly. Under 633-nm laser excitation $I(\nu_2,\bm{c})/I(\nu_3,\bm{c})$ is temperature-independent in all compounds. For d[3,3], $I(\nu_2,\bm{c})/I(\nu_3,\bm{c})$ also depends of the cooling rate. Such temperature dependences and the cooling rate dependence of $I(\nu_2,\bm{c})/I(\nu_3,\bm{c})$ are connected with the fact that $\nu_2$ mode is in resonance with $\bm{c}$-polarised 514-nm laser irradiation but $\nu_3$ is not. At low temperatures resonance conditions improve and $I(\nu_2,\bm{c})$ increases, $I(\nu_3,\bm{c})$ remaining constant. For d[3,3], fast cooling increases disorder and worsens the resonance conditions, acting similar to the heating. 
\begin{figure}[h]
\includegraphics[width=0.5\textwidth,height=0.5\textwidth,keepaspectratio=true]{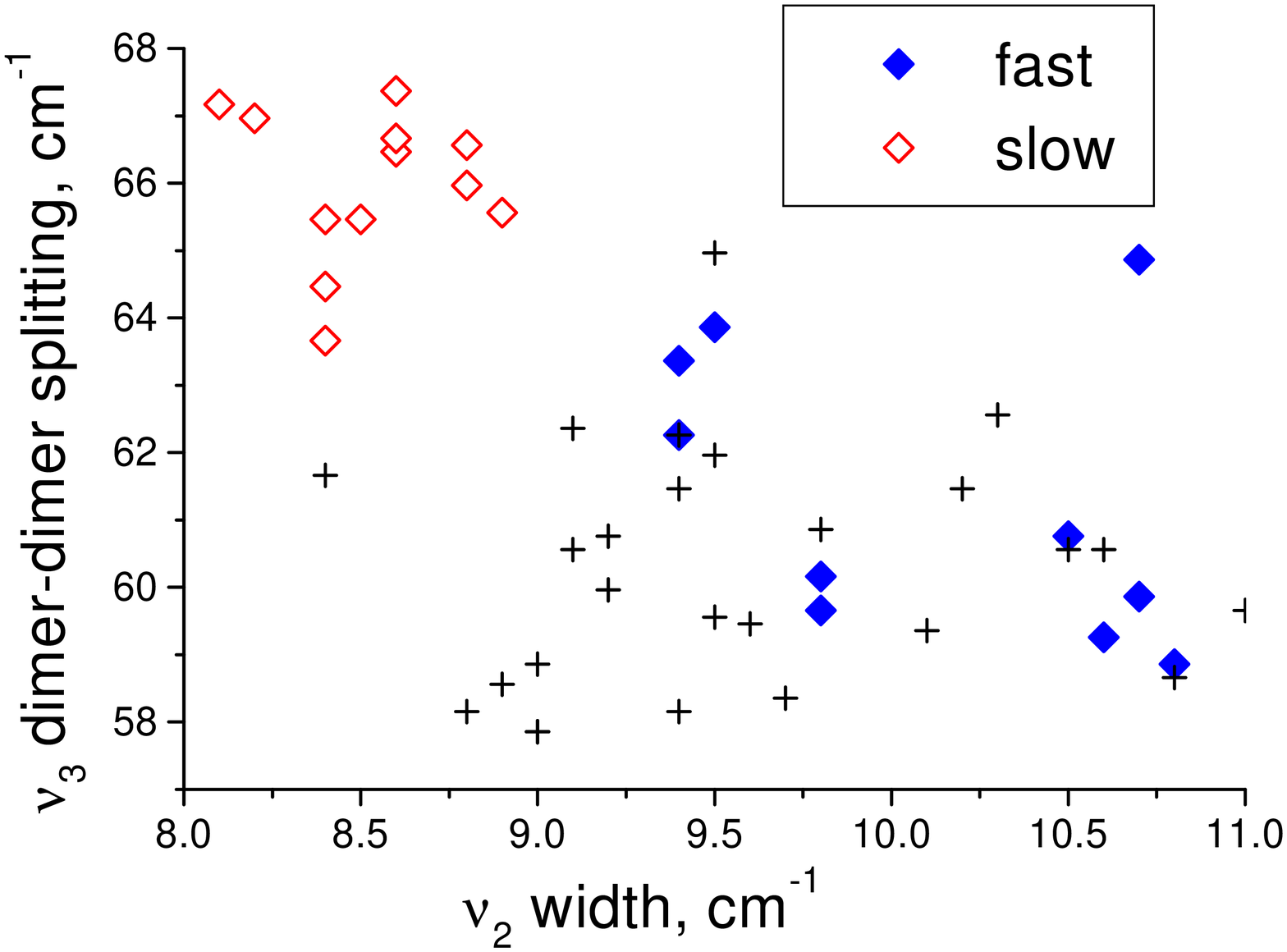}
\caption{\label{fig:fig9} Correlations between $\delta(\nu_2,\bm{c},3)$, and $\omega(\nu_3(A_g+B_{1g}))-\omega(\nu_3(B_{2g}+B_{3g}))$ Temperature is 5 K. Open (\textcolor{red}{$\diamondsuit$}) and solid (\textcolor{blue}{$\blacklozenge$}) diamonds designate the data after "slow" and "fast" cooling  respectively, crosses (+) show the data obtained after different cooling procedures that are neither "slow" nor "fast".}
\end{figure}

Fig.~\ref{fig:fig9} shows the correlations between the parameters influenced by cooling rate for d[3,3]. The temperature was 5 K. We put here also the data obtained in a period when we were looking for a good "slow" cooling procedure and were trying different methods ('+' symbols). In Fig.~\ref{fig:fig9} "slow" and "fast" data occupy  different areas, especially in $\delta(\nu_2)$ direction. Except that, however, there are no evident correlations between $\delta(\nu_2)$ and the dimer--dimer splitting of $\nu_3$. The same can be said about the other possible pairs (not shown): $\delta(\nu_2), I(\nu_2,\bm{c})/I(\nu_3,\bm{c})$; and $I(\nu_2,\bm{c})/I(\nu_3,\bm{c}), \nu_3$ dimer--dimer splitting. 

d[3,3] is very sensitive to a cooling procedure, more than d[2,2]. We found that when we cooled the sample with approximately constant rate 1 K/min from 150 down to 50 K without any interruptions, we obtained the biggest difference with the "fast" cooling procedure. This temperature region can not be narrower. We also tried annealing for 6 hours at 70 K or 30 minutes annealing after every 10 K from 150 down to 50 K but the results were, surprisingly, worse (i. e., the difference with the "fast" cooling was smaller). We cannot explain this. Cooling rate 0.5 K/min makes no difference from 1 K/min. Contrary, near the "fast" cooling limit the cooling rate looks quite important, because stable results near the "fast" cooling limit could be obtained only if the transfer tube is well precooled. 

According to Ref. \onlinecite{kawa2} the amount of dielectric/metallic phase for d[4,4] is sensitive to the cooling rate mostly near 60--90 K, at higher temperatures cooling rate being not so important. From the fact that for our data the cooling rate is important in a wider range of temperature (50--150 K) we can suggest that probably any cooling rate dependence found by us results directly from different amount of disorder, not from change of an amount of metallic/antiferromagnic phase produced by the phase separation.

Specifically, the decrease of dimer--dimer EMV coupling below 80 K for d[4,4], d[3,3] (Fig.~\ref{fig:fig8}) can be either an intrinsic property of the insulating phase or a consequence of disorder induced by deuteration and/or fast cooling. We think that the latter looks more reasonable, mostly because for d[3,3] this decrease depends on the cooling rate.
 
The absence of correlations between different parameters shows that these parameters are not directly connected with each other suggesting, again, some kind of disorder.

However the strange thing about this disorder is that the disorder increases with decreasing temperature down from 50 K when the ethylene groups are already completely frozen. Probably for d[4,4] and d[3,3] fast cooling results in rather "bad" antiferromagnet with very small domains or some other kind of spin disorder. Such unknown disorder can give another explanation (except the scattering on the domain boundaries suggested above) for the observed broadening of $\nu_2$ and $\nu_3$ with decreasing temperature.

\section{Conclusions}

1. Below 20--40 K the width of $\nu_2$ and $\nu_3$ Raman lines can increase with decreasing temperature. 

2. Below 50 K, in the antiferromagnetic d[4,4], dimer--dimer EMV splitting of the mode $\nu_3$ decreases with decreasing temperature.

3. A cooling rate dependence has been observed in d[2,2] and d[3,3] crystals for the linewidths of phonon modes both in Raman and IR spectra, and $\nu_3$ dimer--dimer EMV splitting in Raman spectra. The difference between slow and fast cooling can be seen below 20--50 K and increases with decreasing temperature. For d[3,3] the influence of the cooling rate is essential from  $\approx$150 down to $\approx$50 K. We explain the cooling rate dependences as a result of increasing disorder upon fast cooling.

\bibliography{article2}

\end{document}